\documentclass[twocolumn,prc,showpacs,preprintnumbers,aps,10pt]{revtex4-2}
\usepackage{graphicx,latexsym,amssymb,amsmath,color,multirow,mathrsfs,booktabs,ifpdf,epsfig,dcolumn,bm,longtable}
\usepackage{newtxtext}
\usepackage{newtxmath}
\usepackage{changes}
\usepackage{hyperref}
\hypersetup{colorlinks=true,linkcolor=blue,filecolor=magenta,urlcolor=cyan,citecolor=blue}

\begin{document}
	
\title{Constraining neutron skin impurity in $^{48}\mathrm{Ca}$ and its relevance for the CREX-PREX puzzle}

\author{Phan Nhut Huan} 
\email[Corresponding author:~]{phannhuthuan@duytan.edu.vn}
\affiliation{Institute of Research and Development, Duy Tan University, Da Nang 550000, Vietnam}
\affiliation{School of Engineering and Technology, Duy Tan University, Da Nang 550000, Vietnam}

\begin{abstract}
The impact of Coulomb core polarization on the neutron density distribution of $^{48}\mathrm{Ca}$ is investigated. The Coulomb boundary radius is located before the neutron skin region, leading to proton admixture and establishing neutron skin impurity as an intrinsic feature of the density profile. Using the $(^3\mathrm{He},t)$ isobaric analog state (IAS) reaction at 420 MeV, the impurity-induced modification of the probed neutron distribution is quantified. For a neutron skin of 0.144 fm, consistent with the CREX thin-skin scenario, Coulomb core polarization enhances the probed neutron distribution by approximately 0.052 fm. This value is remarkably close to the difference between the CREX and RCNP neutron skin determinations, providing a physical mechanism to reconcile the discrepancy between electroweak and hadronic probes in $^{48}\mathrm{Ca}$. These findings provide new insights into the CREX-PREX puzzle, highlight the importance of accounting for Coulomb core polarization in neutron distribution analyses, and motivate future IAS charge-exchange measurements as a complementary hadronic probe.
\end{abstract}


\maketitle

The direct detection of gravitational waves from binary neutron star mergers has opened a new era of multi-messenger astronomy, establishing a strong connection between nuclear physics and astrophysics \cite{GW170817,Nunes20,Somasundaram25}. Nuclear physics plays a central role in this context by constraining the equation of state (EOS) of neutron-rich matter and by describing the rapid neutron-capture process responsible for heavy element synthesis \cite{Khan12,Hebeler15,Cowan21}. A key observable linking finite nuclei to neutron stars is the neutron skin thickness, defined as the difference between the root-mean-square (rms) radius of neutron and proton distributions \cite{Brown00,Horowitz01}. This quantity serves as a sensitive probe of the pressure of neutron-rich matter near saturation density and has important implications for neutron star properties, including their radius and tidal deformability \cite{Roca11,Tsang12,Hagen16,Reed21,Essick21,Hu22}.

The Lead Radius Experiment (PREX-II) reported a relatively thick neutron skin for $^{208}\mathrm{Pb}$ of $0.283 \pm 0.071$ fm, implying a large slope of the symmetry energy ($L = 106 \pm 37$ MeV) and a stiff EOS \cite{PREX-II}. From an astrophysical perspective, such a stiff EOS is associated with larger neutron star radius and higher tidal deformability. In contrast, the Calcium Radius Experiment (CREX) reported a significantly thinner neutron skin for $^{48}\mathrm{Ca}$ of $0.121 \pm 0.035$ fm, indicating a softer EOS \cite{CREX}. This apparent inconsistency has led to the CREX-PREX puzzle, as most theoretical models predict a strong correlation between the neutron skins of medium and heavy nuclei, governed by the same bulk properties of nuclear matter \cite{Reinhard21,Reinhard22}.

Recent theoretical attempts to resolve this tension have explored mechanisms beyond conventional nuclear structure descriptions, such as an enhanced isovector spin-orbit interaction, which shifts neutron orbitals to reduce the neutron skin thickness in $^{48}\mathrm{Ca}$ while having only a marginal impact on $^{208}\mathrm{Pb}$ \cite{Kunjipurayil25}. However, the strong enhancement required to mitigate the CREX-PREX puzzle can lead to inconsistencies with established nuclear structure features, including the level ordering of spin-orbit partners and the stability of magic numbers. Meanwhile, theoretical analyses based on energy density functionals (EDFs) indicate a thinner neutron skin of $^{208}\mathrm{Pb}$, approximately $0.19 \pm 0.02$ fm \cite{Reinhard21}, consistent with astrophysical constraints from the NICER mission \cite{Miller19,Miller21} and the tidal deformability extracted from the gravitational wave event GW170817 \cite{Raithel18,Annala18,Lim18}. 

In this context, Coulomb core polarization constitutes an important mechanism that can influence neutron skin observables. The long-range Coulomb repulsion acting on protons drives them toward the nuclear surface, leading to a redistribution of protons in the nuclear interior and their admixture in the neutron-rich outer region \cite{Huan24}. To quantify this effect, the nucleus with $N > Z$ is described as a symmetric core of $Z$ protons and $Z$ neutrons, surrounded by $N-Z$ excess neutrons. Following Ref.~\cite{Auerbach81}, the isovector density is decomposed as $\rho_n(r) - \rho_p(r) = \rho_{n\,\mathrm{exc}}(r) + \delta\rho(r)$, where $\rho_{n\,\mathrm{exc}}(r) = \sum_{i = Z+1}^{N} \left|\varphi_i^n (r)\right|^2$ denotes the density of the $N-Z$ excess neutrons and $\delta \rho(r) = \sum_{i = 1}^{Z} \left|\varphi_i^n (r)\right|^2 - \sum_{i = 1}^{Z} \left|\varphi_i^p (r)\right|^2$ represents the density difference between core neutrons and protons due to Coulomb-driven polarization. Here, $\varphi_i^{n(p)}(r)$ is the single-particle wave function of the neutron (proton). The rms radius of the neutron, proton, and neutron excess distributions are denoted by $R_n$, $R_p$, and $R_{n\,\mathrm{exc}}$, respectively.

The effect of Coulomb core polarization in $^{48}\mathrm{Ca}$ is investigated using Skyrme Hartree-Fock (SHF) calculations with the SAMi-J EDF family \cite{Colo13,Roca13}, chosen for their improved description of spin-isospin properties and consistent mean-field treatment of finite nuclei and nuclear matter \cite{Roca12,Naito23}. By systematically varying the symmetry energy, a set of neutron skin thicknesses is generated that covers the experimental ranges reported by CREX ($0.121 \pm 0.035$ fm)~\cite{CREX} and RCNP ($0.168^{+0.025}_{-0.028}$ fm)~\cite{Zenihiro18}, providing a framework to investigate the dependence of Coulomb core polarization on neutron skin thickness, as summarized in Table~\ref{48Ca}. The corresponding values of $\Delta R_{\mathrm{core}}$ exhibit the same systematic behavior previously reported in Ref.~\cite{Huan24}, becoming less negative as the neutron skin increases. To quantify proton redistribution, the quantity $S = -4\pi \int_{R_s}^{\infty} \left[ \rho_n(r) - \rho_p(r) - \rho_{n\,\mathrm{exc}}(r) \right] r^2 \,dr$ is evaluated, representing the number of protons shifted into the surface region, following Ref.~\cite{Loc17}, where $R_s$ denotes the position of the last node of the core-polarization density. To enable comparison across nuclei with different neutron excess $N-Z$, the ratio $\alpha = S/(N-Z)$ is used, which quantifies the relative strength of Coulomb core polarization. The decrease of both $S$ and $\alpha$ with increasing neutron skin thickness indicates that the relative impact of Coulomb core polarization weakens as the symmetry potential strengthens.

\begin{table}[htb]
	\caption{Properties of $^{48}\mathrm{Ca}$ obtained from Skyrme Hartree--Fock calculations using the SAMi-J interaction. The neutron skin thickness is $\Delta R_{np} = R_n - R_p$, the neutron excess thickness is $\Delta R_{n\,\mathrm{exc}} = R_{n\,\mathrm{exc}} - R_p$, and $\Delta R_{\mathrm{core}} = R_{n\,\mathrm{core}} - R_{p\,\mathrm{core}}$ reflects the impact of Coulomb core polarization, with $R_{p\,\mathrm{core}} = R_p$. $S$ denotes the number of protons driven into the surface region, and $\alpha = S/(N-Z)$ quantifies the relative strength of the Coulomb core polarization effect. All root-mean-square (rms) radius are given in fm.}
	\resizebox{\columnwidth}{!}{
	\begin{tabular}{cccccccccc}
		\hline
		\hline 
		J & $R_n$ & $R_{p}$ & $\Delta R_{np}$  & $R_{n\,\mathrm{core}}$ & $\Delta R_{\mathrm{core}}$ & $R_{n\,\mathrm{exc}}$ & $\Delta R_{n\,\mathrm{exc}}$ & S & $\alpha$  \\
		\hline 
		27 & 3.588 & 3.443 & 0.144 & 3.335 & -0.108 & 4.152 & 0.709 & 1.017 & 0.127 \\
		28 & 3.590 & 3.439 & 0.151 & 3.338 & -0.101 & 4.154 & 0.715 & 0.971 & 0.121 \\
		29 & 3.616 & 3.432 & 0.184 & 3.364 & -0.068 & 4.180 & 0.748 & 0.691 & 0.086 \\
		30 & 3.629 & 3.425 & 0.203 & 3.378 & -0.047 & 4.190 & 0.765 & 0.541 & 0.068 \\
		\hline		
		\hline
		\label{48Ca}
	\end{tabular}
}
\end{table}

\begin{figure}[h!]
	\centering
	\includegraphics[width=0.95\columnwidth]{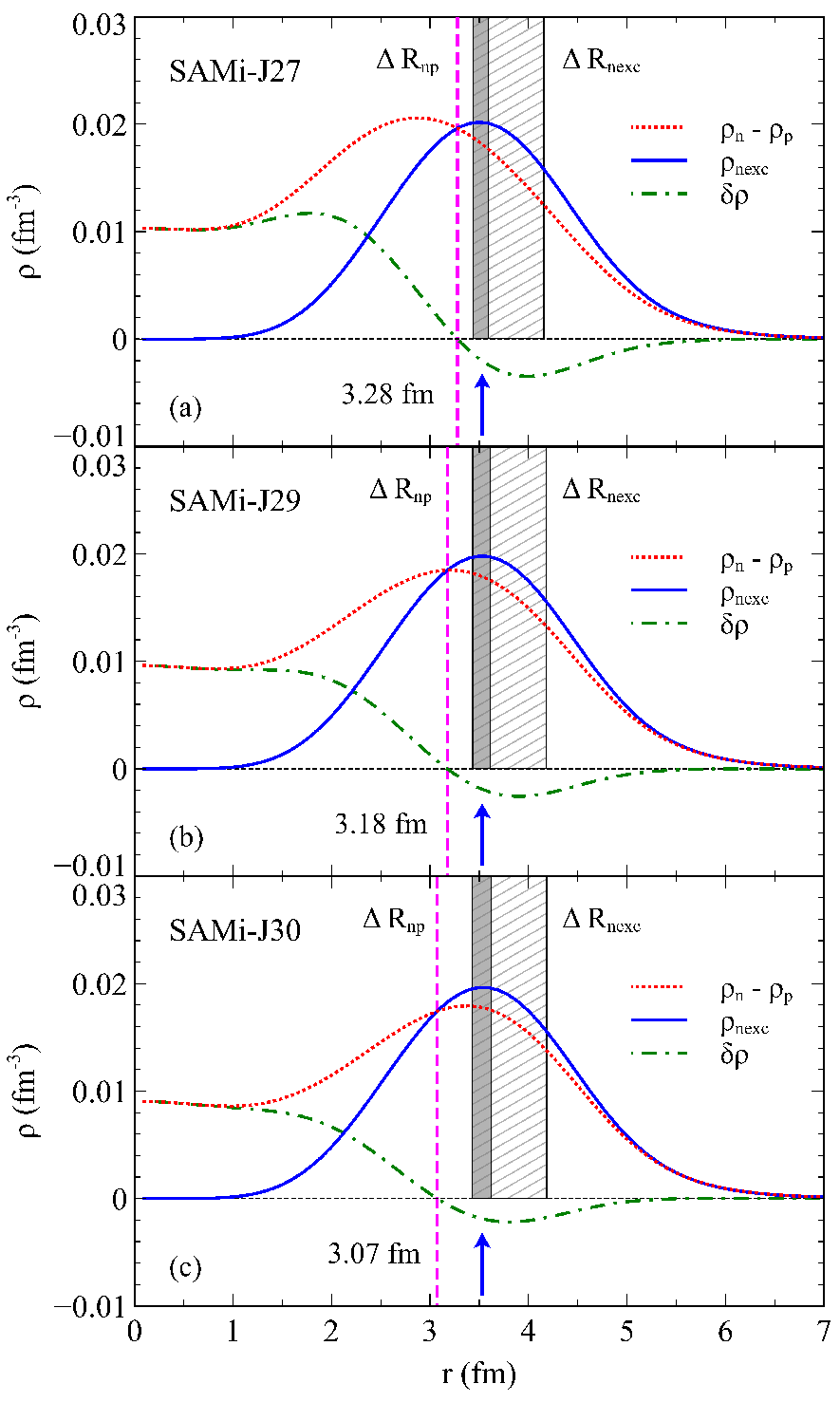}
	\caption{Nuclear density distributions of $^{48}\mathrm{Ca}$ calculated with the SAMi-J27, SAMi-J29, and SAMi-J30 interactions, showing the isovector density $(\rho_n - \rho_p)$, the neutron excess density $(\rho_{n\,\mathrm{exc}})$, and the Coulomb core-polarization density $(\delta\rho)$. The neutron skin thickness ($\Delta R_{np}$) is indicated by the gray area, while the neutron excess thickness ($\Delta R_{n\,\mathrm{exc}}$) is shown by the forward-slashed area. The vertical dashed line indicates the Coulomb boundary radius $R_B$, and the neutron skin impurity phenomenon, where protons are mixed into the neutron skin, is highlighted by the blue arrow.}
	\label{Dens}
\end{figure}

To illustrate this behavior, the density distributions of $^{48}\mathrm{Ca}$ obtained with the SAMi-J27, SAMi-J29, and SAMi-J30 interactions are shown in Fig.~\ref{Dens}. The impact of Coulomb core polarization on neutron distributions involves two aspects. The first aspect is the discrepancy between the neutron excess density $(\rho_{n\,\mathrm{exc}})$ and the conventional isovector density $(\rho_n-\rho_p)$ in the nuclear surface region, arising from the outward redistribution of protons driven by the Coulomb interaction. This effect is reflected in the core polarization contribution $\delta\rho$, which exhibits the characteristic nodal structure previously observed in $^{208}\mathrm{Pb}$~\cite{Huan24}. The second aspect, and more important, is the position of the Coulomb boundary radius $R_B$, defined as the radial position at which protons begin to be driven outward by the Coulomb force. The location of $R_B$ is governed by the competition between the Coulomb repulsion, characterized by the proton number $Z$, which pushes protons toward the nuclear surface, and the symmetry potential, associated with the neutron excess $(N-Z)$, which acts to counterbalance this redistribution. This competition leads to fundamentally different density configurations in $^{208}\mathrm{Pb}$ and $^{48}\mathrm{Ca}$. In $^{208}\mathrm{Pb}$, the large neutron excess ($N-Z=44$) generates a sufficiently strong symmetry potential to maintain the configuration $R_p<R_n<R_B$ for moderate neutron skin thicknesses, keeping $R_B$ beyond the neutron skin region and thereby preventing proton admixture into the neutron skin \cite{Huan24}. In contrast, the much smaller neutron excess in $^{48}\mathrm{Ca}$ ($N-Z=8$) provides a weaker restoring effect, leading to a configuration that satisfies $R_B<R_p<R_n$ throughout the neutron skin range considered in this work. As a result, protons are admixed into the neutron skin region, establishing neutron skin impurity as an intrinsic feature of the density profile in $^{48}\mathrm{Ca}$. The $R_B$ shifts systematically inward from 3.28 fm for SAMi-J27 to 3.18 fm for SAMi-J29 and 3.07 fm for SAMi-J30, reflecting the increasing role of the symmetry potential as the neutron skin thickness increases, which progressively suppresses Coulomb-driven proton redistribution and diminishes the impact of neutron skin impurity.

The distinction between $\rho_{n\,\mathrm{exc}}$ and $\rho_n-\rho_p$ induced by Coulomb core polarization has important implications for the interpretation of surface-sensitive hadronic probes \cite{Huan24,Huan21,Huan23}. While elastic scattering primarily probes the total matter distribution $(\rho_n+\rho_p)$, and extracts neutron distribution indirectly through reaction model analyses, the $(^3\mathrm{He},t)$IAS charge-exchange reaction at intermediate energies is directly sensitive to the isovector distribution $(\rho_n-\rho_p)$, and therefore, can probe the fine structure of the nuclear surface \cite{Zegers07,Loc14}. As a result, the presence of neutron skin impurity in $^{48}\mathrm{Ca}$ can modify the neutron distribution effectively probed by the reaction and influence the extraction of neutron skin properties from cross section analyses. More broadly, hadronic probes have larger uncertainties than electroweak probes due to the complexity of the strong interaction and the unavoidable model dependence involved in reaction analyses \cite{Mammei24}. This underscores the need for continued efforts to refine hadronic determinations of neutron skin thickness. Collectively, these observations suggest that Coulomb core polarization and the resulting neutron skin impurity constitute an intrinsic nuclear structure effect that must be taken into account in such analyses.

The effect of neutron skin impurity on the $(^3\mathrm{He},t)$IAS reaction is investigated within the Lane model \cite{Lane62} and the distorted wave Born approximation (DWBA). The nucleus-nucleus optical potential is determined using a double-folding model \cite{Khoa00,Toyokawa15,Durant18}, employing the Chiral three-nucleon forces (3NFs) $G$-matrix interaction derived from Brueckner-Hartree-Fock (BHF) calculations in nuclear matter \cite{Toyokawa18}. The $^{48}\mathrm{Ca}$ densities are obtained from SHF calculations with the SAMi-J family \cite{Colo13,Roca13}, while the $^{3}\mathrm{He}$ and triton densities are taken from Ref.~\cite{Nielsen01}. This fully microscopic framework has successfully reproduced IAS charge-exchange experimental data without parameter adjustment \cite{Huan21,Huan24}. Two transition densities, $\rho_n-\rho_p$ and $\rho_{n\,\mathrm{exc}}$, are considered to investigate how Coulomb core polarization modifies the neutron distribution probed by the reaction. Further details of the calculation are provided in the Supplemental Material \cite{Suppl}.

\begin{figure}[h!]
	\centering
	\includegraphics[width=0.85\columnwidth]{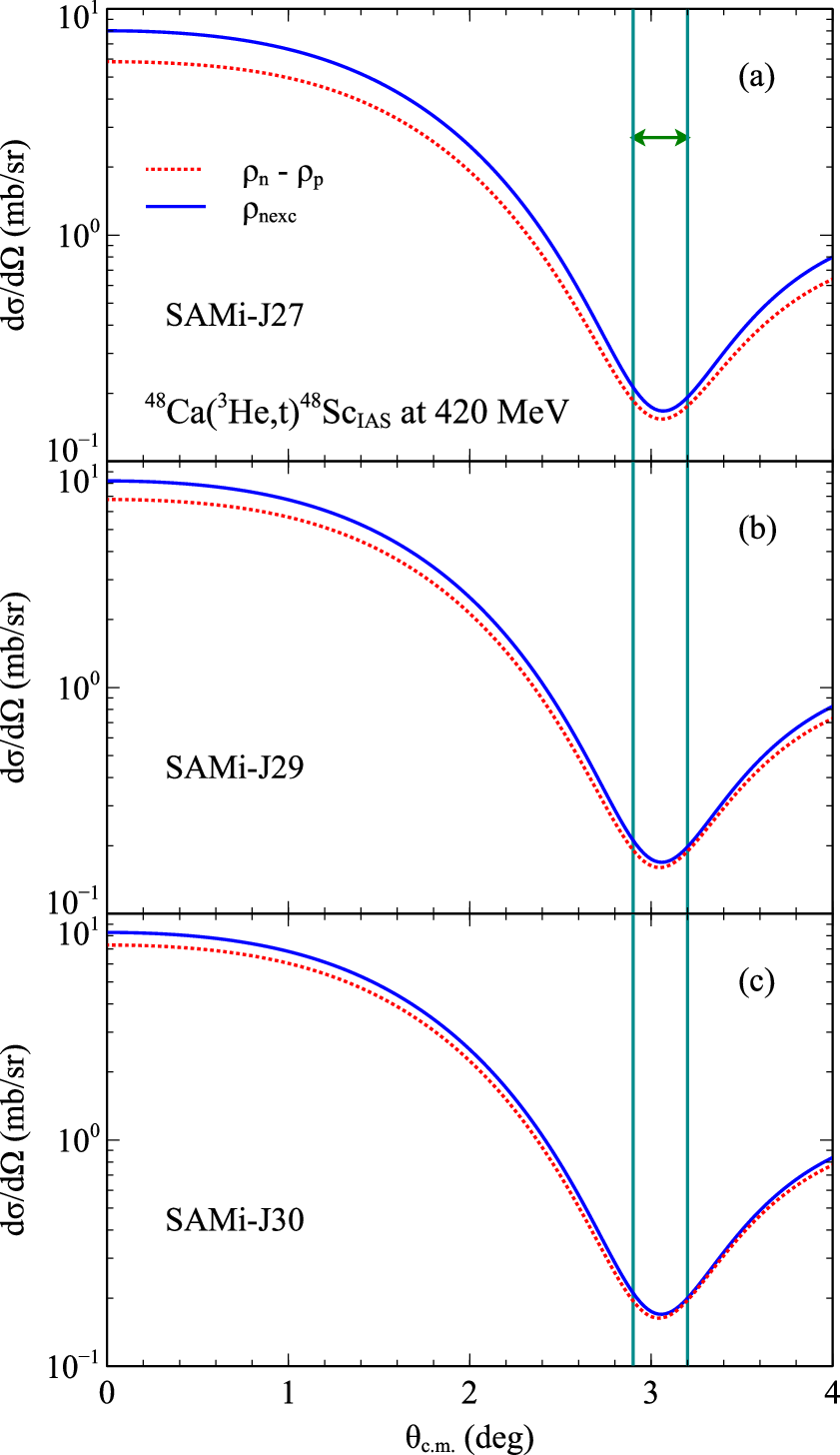}
	\caption{Differential cross sections of the $^{48}\mathrm{Ca}(^{3}\mathrm{He},t)$IAS reaction at 420 MeV calculated using the isovector density $(\rho_n-\rho_p)$ and the neutron excess density $\rho_{n\,\mathrm{exc}}$ for the SAMi-J27 (a), SAMi-J29 (b), and SAMi-J30 (c) interactions. The green horizontal arrow indicates the angular region around the first minimum, where the incident particle probes the impurity-affected region.}
	\label{CS}
\end{figure}

Figure~\ref{CS} presents the differential cross sections of the $^{48}\mathrm{Ca}(^{3}\mathrm{He},t)$IAS reaction at 420 MeV calculated using the isovector density $(\rho_n-\rho_p)$ and the neutron excess density $(\rho_{n\,\mathrm{exc}})$ for the SAMi-J27, SAMi-J29, and SAMi-J30 interactions. At forward scattering angles, calculations using $\rho_{n\,\mathrm{exc}}$ systematically yield larger cross sections than those obtained with $\rho_n-\rho_p$, reflecting the higher density encountered by the incident $^{3}\mathrm{He}$ particle in the nuclear surface region when the neutron excess density is used. This behavior is consistent with that previously observed in $^{208}\mathrm{Pb}$ \cite{Huan24} and represents the first intrinsic source of uncertainty associated with the neutron distribution probed by surface-sensitive hadronic reactions. As the neutron skin thickness increases from SAMi-J27 to SAMi-J30, the difference between the two calculations decreases, indicating a reduced influence of Coulomb core polarization. However, unlike the $^{208}\mathrm{Pb}$ case, neutron skin impurity is intrinsically present in $^{48}\mathrm{Ca}$. Around the first minimum of the angular distribution, the incident particle has already traversed the neutron excess region associated with forward scattering angles and begins to probe the impurity-affected neutron region. As a result, the differential cross section is influenced not only by the discrepancy between $\rho_{n\,\mathrm{exc}}$ and $\rho_n-\rho_p$ but also by the neutron skin impurity induced by Coulomb core polarization, which constitutes a second intrinsic source of uncertainty in the neutron distribution probed by the reaction.

\begin{figure}[h!]
	\centering
	\includegraphics[width=0.95\columnwidth]{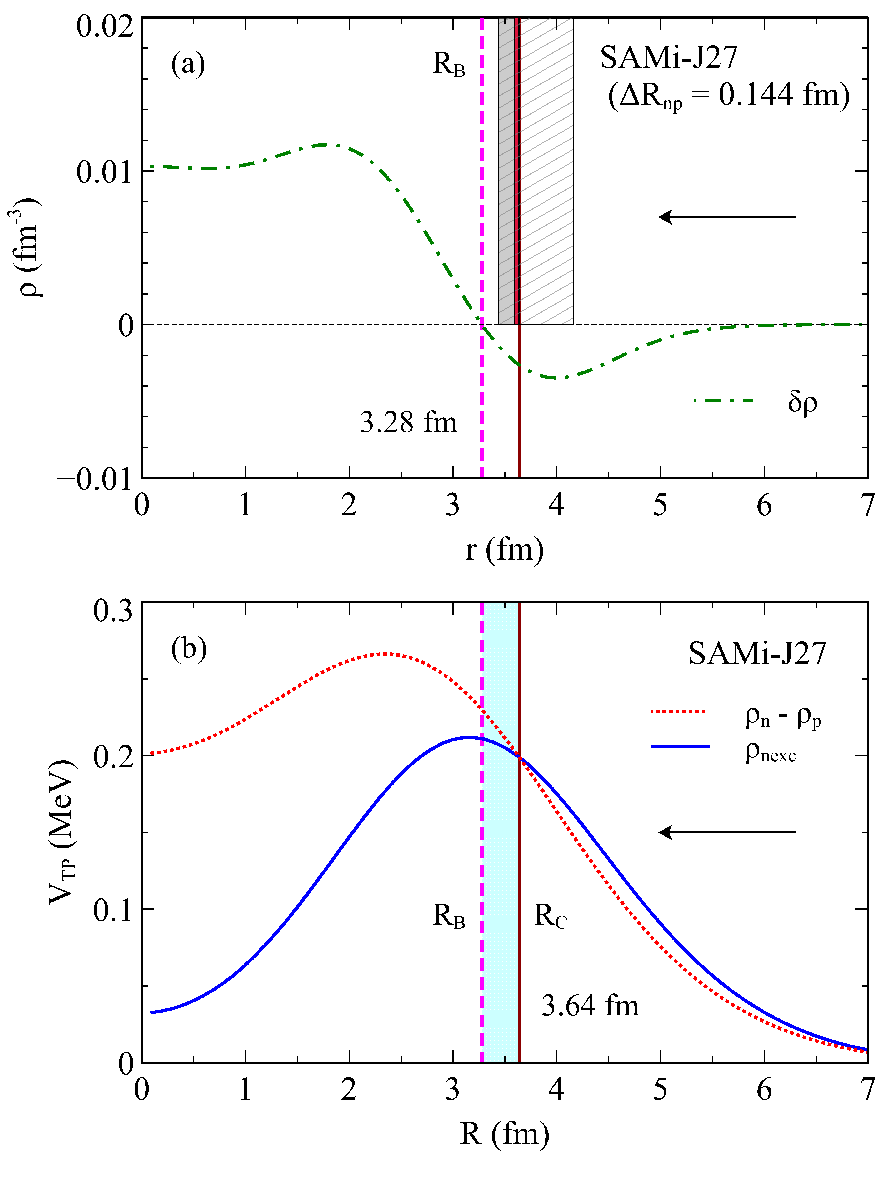}
	\caption{(a) Coulomb core-polarization distribution of $^{48}\mathrm{Ca}$ for the SAMi-J27 interaction. The neutron skin thickness is indicated by the gray region, while the neutron excess thickness is shown by the forward-slashed area. The red region highlights the impurity-induced enhancement $\Delta R_{\mathrm{imp}}$, arising from the mixing of protons into the neutron skin and the resulting extension of the probed neutron distribution. The vertical dashed line indicates the Coulomb boundary radius $R_B$. (b) Real part of the $^{48}\mathrm{Ca}(^{3}\mathrm{He},t)$IAS transition potential calculated using the two choices of transition density. The vertical solid line denotes the convergence point $R_C$, and the horizontal arrow indicates the direction of the incident particle probing the target. The impurity-affected region is highlighted by the blue shaded area.}
	\label{TransImp}
\end{figure}

The primary question is how the incident particle interacts with the impurity-affected neutron region. To explain the mechanism, Fig.~\ref{TransImp} presents the core polarization density distribution $\delta \rho$ and the corresponding transition potentials for the $^{48}\mathrm{Ca}(^{3}\mathrm{He},t)$IAS reaction in the SAMi-J27 case. Figure~\ref{TransImp}(a) illustrates the  $^{48}\mathrm{Ca}$ structure, while Fig.~\ref{TransImp}(b) shows the real part of the transition potential obtained using the two choices of transition density. This component directly reflects the strong interaction between the incident particle and the target nucleus and serves as a critical benchmark for probing the neutron distribution in the IAS reaction. A notable feature is the convergence point $R_C$, defined as the radial position at which the two transition potentials are equal, as indicated by the vertical solid line in Fig.~\ref{TransImp}(b). At this point, both transition density choices yield the same interaction strength between the incident particle and the probed neutron distribution.

The impurity-affected region is defined as the spatial interval between the Coulomb boundary radius ($R_B$) and the convergence point ($R_C$), highlighted by the blue shaded area in Fig.~\ref{TransImp}(b). The existence of this region depends on whether neutron skin impurity is absent or present. In $^{208}\mathrm{Pb}$, the density configuration corresponds to $R_p<R_n<R_B$ for moderate neutron skin thicknesses \cite{Huan24}, placing the Coulomb boundary radius beyond the neutron skin region, and neutron skin impurity is absent. In this case, the only manifestation of Coulomb core polarization is the discrepancy between $\rho_{n\,\mathrm{exc}}$ and $\rho_n-\rho_p$ at the nuclear surface, leading to the coincidence of $R_B$ and $R_C$. In contrast, $^{48}\mathrm{Ca}$ exhibits the configuration $R_B<R_p<R_n$, indicating that protons are admixed into the surface region and that neutron skin impurity becomes an intrinsic feature of the density profile. In this case, $R_B$ and $R_C$ no longer coincide. The physical origin of this separation is the response of the symmetry potential to proton admixture. The long-range Coulomb repulsion drives core protons toward the nuclear surface, locally breaking the isospin balance. To restore isospin equilibrium, the symmetry potential acts as a restoring force, pulling the neutron distribution further outward and thereby modifying the neutron skin probed by the incident particle. As a result, the probed neutron distribution effectively extends beyond $R_B$, generating a finite impurity-affected region between $R_B$ and $R_C$. This mechanism is particularly evident for the SAMi-J27 interaction, which predicts a neutron skin thickness of 0.144 fm, consistent with the thin-skin scenario suggested by CREX, where $R_B=3.28$ fm and $R_C=3.64$ fm, corresponding to an impurity-affected region of approximately 0.36 fm. The resulting impurity-induced enhancement of the neutron distribution is about $\Delta R_{\mathrm{imp}} = R_C - R_n \approx 0.052$ fm, increasing the effective neutron skin thickness probed by the $(^3\mathrm{He},t)$IAS reaction from the intrinsic value of 0.144 fm to approximately 0.196 fm. Importantly, this value is remarkably close to the difference of about 0.047 fm between the neutron skin thicknesses extracted from parity-violating electron scattering (PVES) in CREX, $0.121$ fm \cite{CREX}, and proton elastic scattering at RCNP, $0.168$ fm \cite{Zenihiro18}. This close agreement can be naturally explained by the underlying physical mechanism. Coulomb core polarization and the resulting neutron skin impurity constitute an intrinsic nuclear structure effect that effectively enhances the neutron distribution probed by surface-sensitive hadronic reactions, whereas PVES probes the intrinsic neutron skin more directly because it is largely free from strong interaction uncertainties \cite{Mammei24}. Consequently, neutron skin impurity induced by Coulomb core polarization provides a physical mechanism that can reconcile the discrepancy between electroweak and hadronic determinations of the neutron skin thickness in $^{48}\mathrm{Ca}$.

\begin{figure*}[htb]
	\centering
	\includegraphics[width=0.75\textwidth]{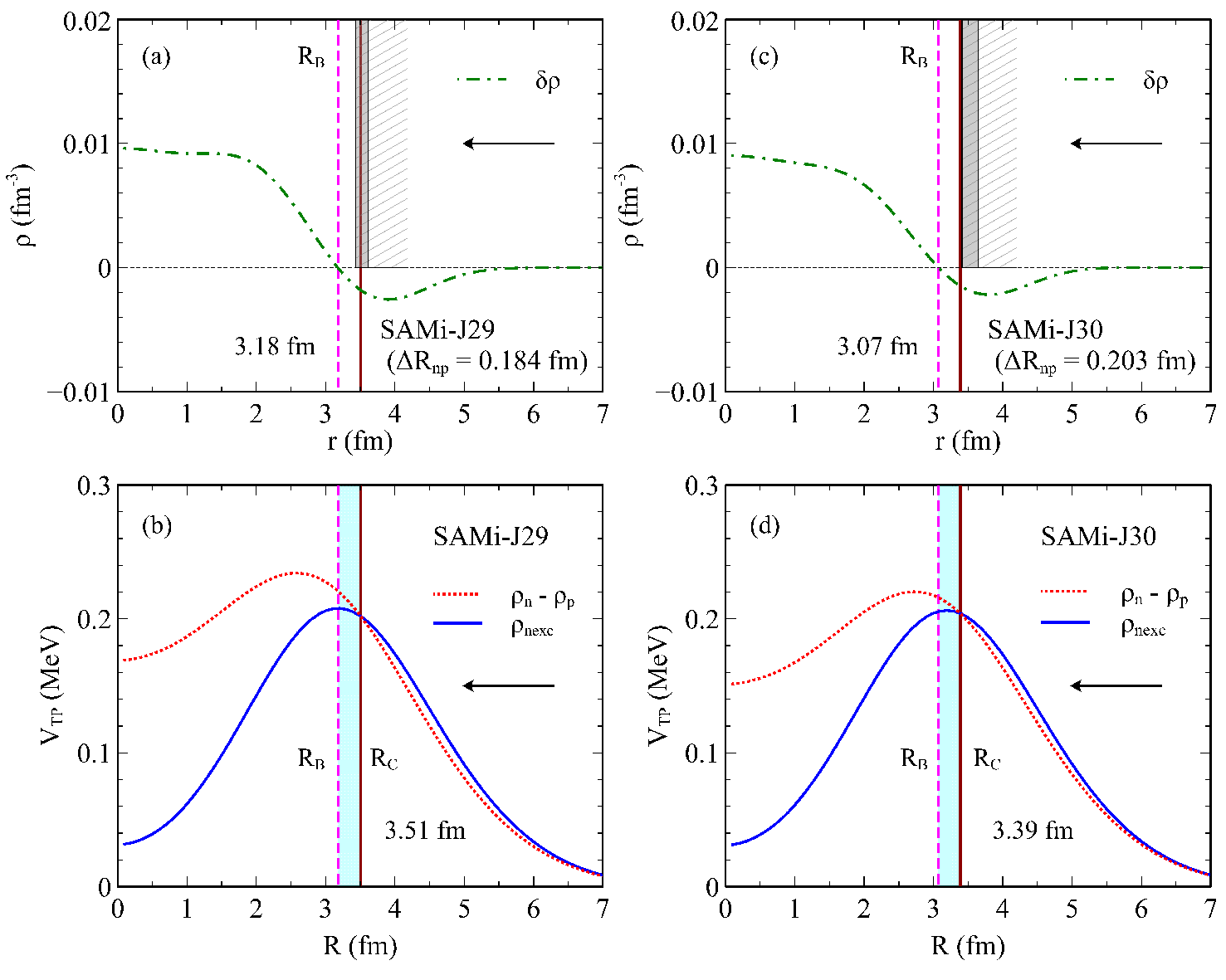}
	\caption{Same as Fig.~\ref{TransImp}, but for the SAMi-J29 interaction in panels (a) and (b), and the SAMi-J30 interaction in panels (c) and (d).}
	\label{TransLargeSkin}
\end{figure*}

As the neutron skin thickness increases, the increasing dominance of the symmetry potential reduces the influence of Coulomb core polarization. For SAMi-J29, shown in Fig.~\ref{TransLargeSkin}(a,b), corresponding to a neutron skin thickness of 0.184 fm, $R_B$ shifts inward to 3.18 fm, while $R_C$ decreases to 3.51 fm, resulting in an impurity-affected region of approximately 0.33 fm. Compared with the SAMi-J27 case, the impurity effect is weakened, as reflected by the reduced core-polarization strength ($S=0.691$ and $\alpha=0.086$). A similar trend is observed for SAMi-J30, shown in Fig.~\ref{TransLargeSkin}(c,d), which predicts a neutron skin thickness of 0.203 fm. In this case, $R_B$ and $R_C$ further decrease to 3.07 fm and 3.39 fm, respectively, yielding an impurity-affected region of approximately 0.32 fm. The corresponding values of $S=0.541$ and $\alpha=0.068$ indicate a further suppression of proton admixture. Notably, $R_C$ becomes smaller than the proton rms radius ($R_C = 3.39~\mathrm{fm} < R_p = 3.425~\mathrm{fm}$), indicating that the symmetry potential effectively counterbalances Coulomb-driven proton redistribution. Consequently, the impurity-affected region is displaced away from the neutron skin domain, strongly suppressing the impurity-induced modification of the neutron distribution and preserving an almost pure neutron skin.

This systematic suppression of the impurity effect with increasing neutron skin establishes a clear connection between Coulomb core polarization and the neutron distribution probed by the $(^3\mathrm{He},t)$IAS reaction. A similar behavior is observed in $^{208}\mathrm{Pb}$ within the SAMi-J EDF framework \cite{Huan24}. For SAMi-J35, which predicts a neutron skin thickness of 0.266 fm consistent with the PREX-II large-skin scenario \cite{PREX-II}, neutron skin impurity exists but is strongly diluted. The values calculated here, $S=1.297$ and $\alpha=0.029$, indicate a much weaker impurity effect than in $^{48}\mathrm{Ca}$, reflecting the stronger symmetry potential due to the large neutron excess in $^{208}\mathrm{Pb}$, which suppresses Coulomb-driven proton redistribution and minimizes proton admixture into the neutron skin region. Consequently, the neutron skin remains nearly pure under the PREX-II constraint. These results are directly relevant to the CREX-PREX puzzle. In $^{48}\mathrm{Ca}$, neutron skin impurity emerges as an intrinsic feature of the density profile and is most pronounced for the thin-skin scenario suggested by CREX, whereas in $^{208}\mathrm{Pb}$ the impurity effect is strongly diluted in the large-skin PREX-II scenario. This contrast highlights the different interplay between the Coulomb repulsion and the symmetry potential in the two systems, demonstrating that Coulomb core polarization introduces an intrinsic nuclear structure mechanism that should be considered when interpreting neutron skin thicknesses extracted from different experimental probes. Accordingly, the present results provide an additional physical ingredient for the broader understanding of the CREX-PREX dilemma.

In summary, this Letter demonstrates that neutron skin impurity is an intrinsic feature of the density profile in $^{48}\mathrm{Ca}$ and becomes most pronounced under the thin-skin constraint provided by CREX. Coulomb core polarization induces a quantitative enhancement of approximately 0.052 fm in the neutron distribution probed by the $(^3\mathrm{He},t)$IAS reaction, offering a physical interpretation of the discrepancy between electroweak and hadronic determinations of the neutron skin in $^{48}\mathrm{Ca}$. A dedicated $^{48}\mathrm{Ca}(^{3}\mathrm{He},t)$IAS measurement at 420 MeV would strongly constrain impurity effects and refine hadronic neutron skin determinations. More broadly, this approach is especially relevant for short-lived neutron-rich nuclei, including doubly magic systems such as $^{132}\mathrm{Sn}$, where electroweak measurements are currently impractical, whereas IAS charge-exchange reactions can be performed in inverse kinematics at next-generation radioactive beam facilities such as FRIB.

\bibliographystyle{apsrev4-2-author-truncate}
\bibliography{Refs}

\end{document}